\documentstyle[preprint,aps]{revtex}

\begin{document}
\draft
\tighten

\preprint{\vbox{\hbox{IMPERIAL/TP/94-95/9}
\hbox{TUTP-95-1}
\hbox{NI94038}
\hbox{hep-ph/9501207}}}

\title{Density of strings formed at a second-order
cosmological phase transition}

\author{T.W.B. Kibble}
\address{Blackett Laboratory, Imperial College, London SW7 2BZ,
United Kingdom and\\
Isaac Newton Institute for Mathematical Sciences, Cambridge
CB3 0EH, United Kingdom}
\author{Alexander Vilenkin}
\address{Institute of Cosmology,
Department of Physics and Astronomy, \\
Tufts University, Medford MA 02155, USA and\\
Isaac Newton Institute for Mathematical Sciences, Cambridge
CB3 0EH, United Kingdom}

\date{}

\maketitle

\begin{abstract}
We discuss the problem of estimating the characteristic length
scale $\xi_{\rm s}$, and hence the initial density, of a system
of cosmic strings formed at a continuous, second-order phase
transition in the early universe.  In particular, we examine the
roles of the correlation length (or string width) $\xi_{\rm c}$
and the Ginzburg length $\xi_{\rm G}$ which defines the
``fuzziness'' of long strings.  We argue that strings acquire a
clear identity only once $\xi_{\rm s}$ exceeds both $\xi_{\rm
c}$ and $\xi_{\rm G}$, and estimate its magnitude at that time.
\end{abstract}


\narrowtext

\section{Introduction}

Strings or vortices may be formed at phase transitions in
condensed-matter systems and in the early universe \cite{str}.
An important problem in either case is to estimate the initial
vortex density just after the transition.  It has recently been
recognized that previously accepted ideas about this are wrong
in important ways, so a re-examination of the question is
timely.  In the present note, we shall confine our attention to
the case of a continuous, second-order transition, involving the
breaking of a global Abelian U(1) symmetry.  (We believe the
results would be similar in the case of a local, gauge symmetry.)

For many purposes, at least in cosmology, the precise nature of
the initial conditions is not important.  The strings apparently
evolve towards a scaling solution that is largely independent of
the initial conditions.  Nevertheless, for some applications,
for example in discussions of string-mediated baryogenesis
\cite{baryo}, and certainly in the condensed-matter field, it is
important.  In any case, it is important to know that the
initial defect density is not essentially zero.

The evolution of cosmic strings has been extensively studied by
numerical simulation.  The initial conditions for these studies
have been set by choosing random phases on some lattice of
points and using the ``geodesic rule'' to decide whether a
string passes through each plaquette \cite{geodes}.  Along each
edge of the lattice the phase is supposed to interpolate between
the two values at its ends by the shorter of the two possible
paths.  A string passes through a plaquette if the net phase
change around it is $\pm2\pi$, rather than 0.  The simplest
implementation of this idea uses a lattice with tetrahedral
cells, so that at most one string can pass into and out of each
cell and there is no ambiguity about how they are connected.

With this algorithm, the lattice spacing determines the initial
string density. It has usually been assumed that it should be
identified with the correlation length $\xi$ of the scalar Higgs
or order-parameter field at the time of string formation.  Then
the initial string density (length per unit volume) is $1/\xi^2$
times a factor of say 1/4 representing the probability of
finding a string in any given cell.  However, an immediate
question arises: what is the ``time of formation''?
Equivalently, at what temperature should we evaluate $\xi$?  The
traditional answer to this question \cite{trad} has been to
choose the Ginzburg temperature, $T_{\rm G}$, which is defined
so that above $T_{\rm G}$, thermal fluctuations from the
broken-symmetry state back over the central hump of the
potential on the length scale $\xi$ are frequent, while below
it, they are rare.  In transitions in weakly coupled theories,
$T_{\rm G}$ is only slightly below the critical temperature
$T_{\rm c}$.

For several reasons, however, this cannot be the whole story.
Firstly, it takes no account of the rate at which the transition
proceeds.  In thermal equilibrium, $\xi$ diverges at the
transition temperature, but if the transition proceeds at any
finite rate any physically defined correlation length must
remain finite, so it is far from obvious that the thermal
equilibrium value of $\xi$ is the right thing to use.  To decide
this question, we have to look at comparative rates.  Secondly,
the argument for using the Ginzburg temperature was that above
$T_{\rm G}$, because of the thermal fluctuations strings do not
have any sort of permanent identity; one should not try to count
them until that point has been reached.  However, the thermal
fluctuations mostly represent the transient appearance and
disappearance of small loops of string, and are not necessarily
relevant to the formation of long strings.

But perhaps the most cogent argument for re-thinking the
criterion comes from experiment.  Zurek \cite{zurek} suggested
that a good test of the cosmic-string formation scenario might
come from studies of the lambda transition in liquid helium.
Hendry et al.\ \cite{helium} have recently performed such an
experiment, using a pressure quench to take the system rapidly
through the transition from normal to superfluid.  The results
certainly seem consistent with the picture of vortices evolving
towards a scaling regime.  (A similar picture has emerged from
studies of the isotropic-to-nematic transition in liquid
crystals \cite{nematic}.)   In liquid helium, however, the
Ginzburg temperature is quite far below the critical
temperature.  In fact, these experiments never reach the regime
below $T_{\rm G}$, so quite clearly it makes no sense to say
that the initial scale of the vortices is determined by the
correlation length at $T_{\rm G}$.

In what follows we shall examine the various length scales in the
problem and how they evolve with time.  This will lead us to a
new formulation of the criterion for estimating the string
density.

\section{The Ginzburg length}

It is useful to have a specific model in mind.  We shall assume
that the theory can be adequately described by a Landau-Ginzburg
model, involving a scalar field $\phi$.  The free-energy density
is assumed to be of the form
 \begin{equation}
{\cal F} = |\dot\phi|^2 + |\nabla\phi|^2 -
\case{1}{2}m^2(T)|\phi|^2 + \case{1}{8}\lambda|\phi|^4.
 \end{equation}
(We use relativistic normalization, but it would be easy to
change to non-relativistic conventions.)  The coefficient of
$|\phi|^2$ vanishes at the critical temperature $T_{\rm c}$ and
for $T<T_{\rm c}$ but not too far below has the form
 \begin{equation}
m^2(T) = m_0^2\left(1-{T\over T_{\rm c}}\right),
 \end{equation}
where $m_0^2\sim\lambda T_{\rm c}^2$.  In the case of liquid
helium, the Landau-Ginzburg model is known to be rather
inaccurate, but it should still give a qualitatively reasonable
description.  We are not at the moment seeking anything better.

In thermal equilibrium at $T<T_{\rm c}$,  $\phi$ acquires a
non-zero mean value, given by
 \begin{equation}
|\phi|^2 = {2m^2(T)/\lambda} \equiv \phi_{\rm eq}^2(T).
 \end{equation}
The thermal-equilibrium correlation length $\xi_{\rm c}$ of
$\phi$ is simply
 \begin{equation}
\xi_{\rm c}(T) = m^{-1}(T) \sim [\lambda T_{\rm c}(T_{\rm
c}-T)]^{-1/2}. \label{xic}
 \end{equation}
This length scale defines the effective width of a string at
temperature $T$.

The problem we want to address is the following.  Given that the
system goes through the phase transition at some known rate, how
do we calculate the initial length scale $\xi_{\rm s}$ of the
string network, and at what temperature $T_{\rm s}$ is this
first possible?  Here $\xi_{\rm s}$ is defined so that the
string density at $T_{\rm s}$ is $1/\xi_{\rm s}^2$.  The idea is
that we can then use the knowledge of $\xi_{\rm s}$ as initial
data for numerical or analytic studies of string evolution.

It is also convenient to introduce another length scale, which we
shall call the {\it Ginzburg length}, $\xi_{\rm G}$.  (The
relation to the Ginzburg temperature will become clear
shortly.)  The Ginzburg length is defined as the largest length
scale on which thermal fluctuations from $\phi_{\rm eq}$ back to
$\phi=0$ are probable.  As we shall see, it is not necessarily
correct to assume thermal equilibrium, but if we do the
condition is simply
 \begin{equation}
\xi_{\rm G}^3\delta{\cal F} = T,
 \end{equation}
where
 \begin{equation}
\delta{\cal F} = {\cal F}(\phi=0)-{\cal F}(\phi_{\rm eq})
= {m^4(T)\over2\lambda}.
 \end{equation}
Assuming that $\lambda\ll1$, this yields
 \begin{equation}
\xi_{\rm G} \sim (\lambda T_{\rm c})^{-1/3}(T_{\rm c}-T)^{-2/3}.
\label{xiG}
 \end{equation}

The Ginzburg length defines the ``fuzziness'' of the strings.
Small loops of size up to $\xi_{\rm G}$ will continually appear
and disappear.  Long strings too will be subject to rapid
fluctuations.  Small fluctuating loops appearing near a long
string may connect to it, causing it to wander in a random
fashion.  So it is difficult to say exactly where a long string
is, within a distance $\xi_{\rm G}$.  Thermal fluctuations are
liable to cause two long strings to intercommute if they come
within a distance $\xi_{\rm G}$ of each other.

The {\it Ginzburg temperature\/}, $T_{\rm G}$ is simply the
temperature at which $\xi_{\rm G}=\xi_{\rm c}$.  From
(\ref{xic}) and (\ref{xiG}), we see that this requires
 \begin{equation}
T_{\rm c}-T_{\rm G} \sim \lambda T_{\rm c},
 \end{equation}
a well-known result.  Note that $\xi_{\rm G}$ decreases faster
than $\xi_{\rm c}$: for temperatures in the range $T_{\rm
G}<T<T_{\rm c}$, $\xi_{\rm G}>\xi_{\rm c}$.

The assumption that thermal equilibrium is maintained is clearly
not entirely correct.  What we are interested in is the
probability of a fluctuation from $\phi_{\rm eq}$ to $\phi=0$ on
some given scale $L$.  The thermal equilibrium properties of the
system, such as $\xi_{\rm c}$ and $\xi_{\rm G}$, are changing on
a time scale of order $t-t_{\rm c}$, where $t_{\rm c}$ is the
time of the phase transition.  So the thermal equilibrium
calculation above is likely to be a good estimate provided that
$\xi_{\rm G} < t-t_{\rm c}$. In the cosmological case, we have
$T^2t\sim M_{\rm P}$, where $M_{\rm P}$ is the Planck mass.
This inequality then yields
 \begin{equation}
{T_{\rm c}-T\over T_{\rm c}} > {1\over\lambda^{1/5}}
\left(T_{\rm c}\over M_{\rm P}\right)^{3/5}.
\label{ConsCond}
 \end{equation}
We conclude that (\ref{xiG}) should be reasonably reliable except
{\it very\/} close to the transition.

For a condensed-matter system cooled through its transition
temperature at a given rate $dT/dt$, the equivalent condition
would be $\xi_{\rm G} < v(t-t_{\rm c})$, where $v$ is an
appropriate characteristic speed, say the speed of second
sound.  This would give
 \begin{equation}
{T_{\rm c}-T\over T_{\rm c}} > {1\over\lambda^{1/5}}
\left({1\over vT_{\rm c}^2}\left| dT\over dt
\right|\right)^{3/5}.
 \end{equation}
However, this is not
directly applicable to the liquid-helium experiments, because
temperature is not in fact the controlling variable.  It would
be more accurate to say that the transition proceeds because the
sudden decrease of pressure leads to an increase of $T_{\rm c}$.

\section{The string length scale}

If the system is maintained in thermal equilibrium at a
temperature below $T_{\rm G}$, almost all strings will
eventually disappear.  The equilibrium abundance of loops of
length $L$ is determined by the Boltzmann factor
$\exp[-\mu(T)L/T]$, where $\mu(T)\sim m^2(T)/\lambda$ is the
energy per unit length of strings at temperature $T$.  Thus the
loop density at temperature $T$ is exponentially suppressed for
loop sizes greater than
 \begin{equation}
L \sim {\lambda T\over m^2(T)} \sim {1\over T_{\rm c}-T}.
 \end{equation}
For temperatures below the Ginzburg temperature, one finds that
$L$ is less than the string width.

At temperatures in the range $T_{\rm G}<T<T_{\rm c}$, there may
be a significant equilibrium distribution of small loops, but no
long strings would survive once thermal equilibrium had been
reached.  The crucial point, however, is that it takes a very
long time (compared to the time required to establish thermal
equilibrium in other respects) for long strings to disappear.
Qualitatively, this is the reason for the appearance of a
scaling solution.  The time required for string structures on a
given scale $L$ to disappear increases with $L$, so that after
some time only the structures with large $L$ survive.

If we look at the system shortly after the phasetransition, when
the temperature is still above $T_{\rm G}$, we shall see violent
fluctuations on small scales. As we noted above, it is difficult
to say exactly where a long string is, to within a distance
$\xi_{\rm G}$.  Nevertheless, viewed on a larger scale, it can
be seen to have a more permanent existence.  Long strings may
indeed wander on a short time scale, but no small-scale
fluctuation can make them disappear.  Their large-scale
configuration can change only if they encounter another long
string and exchange partners with it.  The long strings cannot
be identified individually until their mean separation,
$\xi_{\rm s}$ say, exceeds $\xi_{\rm G}$.

How then can we estimate the string density, excluding the small
transient loops?  Ideally, we would like to be able to follow
the process dynamically, starting from an initial state in the
symmetric phase above the transition, but that is very hard to
do.  We adopt instead a more indirect approach, in which the
relevant length scale is computed by a self-consistency argument.

We suppose that there is a temperature $T_{\rm s}$, which
we call the string-formation temperature, below which it is
possible to identify strings, and try to estimate $T_{\rm s}$ by
following the {\it subsequent\/} evolution of the strings.  Of
course, $T_{\rm s}$ must depend on the rate at which the
transition proceeds, {\it i.e.}, on the expansion rate of the
Universe.

Shortly after string formation, the strings are moving in a dense
environment and are heavily damped.  The force per unit length
on a string moving with velocity $v$ through this environment
will be given by
 \begin{equation}
f \sim n\sigma v_T \Delta p,
 \end{equation}
where $n$ is the particle density, $\sigma$ the linear cross
section for string-particle scattering, $v_T$ the thermal
velocity and $\Delta p$ the mean momentum transfer.  In the
familiar case of a thin string, we usually take $n\sim T^3$,
$\sigma\sim T^{-1}$, $v_T\sim1$ and $\Delta p\sim vT$, so that
$f\sim T^3v$.  In our case, however, the strings are still very
thick, in the sense that $\xi_{\rm c}=m^{-1}\gg T^{-1}$.  Thus
particles with thermal wavelengths are not scattered by the
string, but simply pass through it unaffected.  We should
therefore include only particles with wavelengths comparable to
or larger than the string width, {\it i.e.}, with momenta
$k<\xi_{\rm c}^{-1}$.  Their density is $n\sim T/\xi_{\rm c}^2$,
and with $\sigma\sim\xi_{\rm c}$, $\Delta p\sim v/\xi_{\rm c}$,
we get
 \begin{equation}
f \sim vTm^2(T).
 \end{equation}

If the length scale of the long string configuration is $\xi_{\rm
s}$, then this damping force must be matched by the force due to
the string tension, which on average is $\mu(T)/\xi_{\rm s}$,
where $\mu(T)\sim m^2(T)/\lambda$, so the typical string
velocity will be
 \begin{equation}
v \sim 1/(\lambda T_{\rm c}\xi_{\rm s}).
 \end{equation}
The characteristic time on which the length scale $\xi_{\rm s}$
will grow will be the time it takes for the string to move a
distance $\xi_{\rm s}$, namely $\xi_{\rm s}/v$.  Thus we expect
 \begin{equation}
{d\xi_{\rm s}\over dt} \sim {v\over\xi_{\rm s}}\xi_{\rm s}
\sim {1\over\lambda T_{\rm c}\xi_{\rm s}}.
 \end{equation}
So, unless $\xi_{\rm s}$ is initially very large, it will tend to
grow as $(t-t_{\rm c})^{1/2}$.  (This is also the growth law
that has been deduced for the defects formed at phase
transitions in various condensed-matter systems \cite{bray}.)
We can estimate the value of $\xi_{\rm s}$ as
 \begin{equation}
\xi_{\rm s} \sim \left(t-t_{\rm c}\over
\lambda T_{\rm c}\right)^{1/2}.
 \end{equation}

In the cosmological case, we can use the time-temperature
relation to write this as
 \begin{equation}
\xi_{\rm s} \sim \left(M_{\rm P}(T_{\rm c}-T)
\over\lambda T_{\rm c}^4\right)^{1/2}.
 \end{equation}

This calculation does {\it not\/} apply very close to $T_{\rm
c}$.  In that region it is not possible to identify individual
long strings, though one can use the methods of high-temperature
field theory to follow the evolution of the string density, if
strings are interpreted simply as loci of zeros of the Higgs
field \cite{ray}.  In the immediate vicinity of $T_{\rm c}$ the
separation between zeros is typically $\xi_{\rm s}\sim T_{\rm
c}^{-1}$.  It then probably grows like some power of $t-t_{\rm
c}$ until we reach the range where the above calculation can be
used.

We can now compare this length with the other length scales to
see when the argument can be used.  Clearly, the discussion only
makes sense if the length scale of the long strings is larger
than their width, {\it i.e.}, $\xi_{\rm s}>\xi_{\rm c}$.  But in
addition, we cannot really identify individual long strings
until their length scale is larger than the Ginzburg length, so
we also require that $\xi_{\rm s}>\xi_{\rm G}$.  We can
reasonably take the string-formation temperature to be the point
at which both these inequalities are first satisfied.  It is
easy to check that
 \begin{equation}
\xi_{\rm s}=\xi_{\rm c} \iff
{T_{\rm c}-T\over T_{\rm c}} \sim
\left(T_{\rm c}\over M_{\rm P}\right)^{1/2}.
\label{scEq}
 \end{equation}
whereas
 \begin{equation}
\xi_{\rm s}=\xi_{\rm G} \iff
{T_{\rm c}-T\over T_{\rm c}} \sim \lambda^{1/7}
\left(T_{\rm c}\over M_{\rm P}\right)^{3/7}.
\label{sGEq}
 \end{equation}

We are interested in the equality which occurs later, or at a
lower value of $T$.  Thus the relevant condition is (\ref{sGEq})
rather than (\ref{scEq}) provided that
 \begin{equation}
T_{\rm c}/M_{\rm P} < \lambda^2.
\label{Glater}
 \end{equation}
Except in cases when $\lambda$ is very small, this condition will
be satisfied for most of the relevant transitions.  For
GUT-scale transitions, it may be marginal, while for transitions
at lower scales it will almost always hold.  In any case, we are
{\it never\/} likely to have a situation where $T_{\rm c}/M_{\rm
P}\gg\lambda$.

If $T_{\rm s}$ is indeed given by (\ref{sGEq}), then the
corresponding length scale at the time is
 \begin{equation}
\xi_{\rm s} \sim \xi_{\rm G} \sim {1\over\lambda^{3/7}}
\left(M_{\rm P}\over T_{\rm c}\right)^{2/7}{1\over T_{\rm c}}.
\label{xifin}
 \end{equation}

We can also now check the consistency of our calculation of
$\xi_{\rm G}$ in the relevant temperature range.  The required
condition was (\ref{ConsCond}). It is easy to check that this
will be satisfied at the temperature $T_{\rm s}$ given by
(\ref{sGEq}) if and only if the condition (\ref{Glater}) is
satisfied.  Thus when $\xi_{\rm G}$ is the relevant scale
length, our calculation of it was valid.

\section{Conclusions}

There are several important length scales in this problem: the
correlation length $\xi_{\rm c}$ which determines the width of
strings; the Ginzburg length $\xi_{\rm G}$ which characterizes
the extent of their ``fuzziness'' due to thermal fluctuations,
and the typical separation $\xi_{\rm s}$ between the strings.
Just after the phase transition, $\xi_{\rm G}>\xi_{\rm s}$; both
are decreasing but $\xi_{\rm G}$ falls faster and eventually the
two become equal at the Ginzburg temperature $T_{\rm G}$.

The typical separation $\xi_{\rm s}$ evolves dynamically,
increasing from an initial value $\sim T_{\rm c}^{-1}$ at the
transition.  Individual strings can only be identified once
$\xi_{\rm s}$ becomes larger than both $\xi_{\rm c}$ and
$\xi_{\rm G}$.  By looking at the later evolution of $\xi_{\rm
s}$, we have estimated the temperature $T_{\rm s}$ at which this
occurs.  In almost all cases, the relevant criterion is
(\ref{sGEq}).  The corresponding length scale is (\ref{xifin}).
This essentially completes our task.

It would be interesting to apply a similar analysis to the case
of the evolution following a pressure-induced quench in liquid
helium.  In that case, there could be interesting phenomena
dependent on parameters such as the initial temperature from
which the quench starts.  We expect in particular that the
statistical properties of the string network can be
substantially modified if the quench starts very close to $T_c$.

\acknowledgments
We are grateful for the hospitality of the Isaac Newton Institute
for Mathematical Sciences, Cambridge, England, where this work
was started.  We have benefited from discussions with many
colleagues, in particular Wojtek Zurek, Tanmay Vachaspati,
Julian Borrill, Ray Rivers and Alasdair Gill. The work of AV was
supported in part by the National Science Foundation.

 \end{document}